\documentclass[12pt]{article}
\usepackage{axodraw}

\catcode`@=12
\begin{document}
\thispagestyle{empty}
\begin{flushright} 
UCRHEP-T385\\ 
February 2005\
\end{flushright}
\vspace{0.5in}
\begin{center}
{\LARGE	\bf Triplicity of Quarks and Leptons\\}
\vspace{1.5in}
{\bf Ernest Ma\\}
\vspace{0.2in}
{\sl Physics Department, University of California, Riverside, 
California 92521, USA\\}
\vspace{1.5in}
\end{center}

\begin{abstract}\
Quarks come in three colors and have electric charges in multiples of 
one-third.  There are also three families of quarks and leptons.  Whereas 
the first two properties can be understood in terms of unification symmetries 
such as $SU(5)$, $SO(10)$, or $E_6$, why there should only be three families 
remains a mystery.  I propose how all three properties involving the number 
three are connected in a fivefold application of the gauge symmetry $SU(3)$.
\end{abstract}

\newpage
\baselineskip 24pt

The fundamental building blocks of particle physics are quarks and leptons. 
The former have electric charges of 2/3 and $-1/3$, and are triplets under 
the unbroken gauge symmetry $SU(3)_C$ of Quantum Chromodynamics.  The latter 
have electric charges 0 and $-1$, and do not have $SU(3)_C$ interactions.  
There are also 3 families of quarks and leptons, each one transforming in 
the same way under the standard $SU(3)_C \times SU(2)_L \times U(1)_Y$ 
gauge group.  The ubiquitous occurrence of the number three may be 
indicative of an underlying symmetry larger than that of the present 
observed Standard Model.  Although each family of quarks and leptons may 
be considered as components of ${\bf 5}^*$ + {\bf 10} under $SU(5)$, or 
of {\bf 16} under $SO(10)$, or of {\bf 27} under $E_6$, the existence 
of 3 families remains unexplained in this context.

A strong hint as to what the underlying symmetry could be comes from the 
maximal subgroup of $E_6$, i.e. $SU(3)_C \times SU(3)_L \times SU(3)_R$, 
under which quarks are contained in the representations $(3,3^*,1)$ and 
$(3^*,1,3)$, and leptons in $(1,3,3^*)$.  Here I propose the following 
extension.  Let the gauge symmetry be $SU(3)_C \times SU(3)_L 
\times SU(3)_M \times SU(3)_R \times SU(3)_F$,  where $SU(3)_F$ is the 
family symmetry and $SU(3)_M$ is the missing link in the lepton sector 
which allows this scheme to work.  The gauge symmetry $SU(3)_F$ is assumed 
to be broken at or above the unification scale already to its non-Abelian 
discrete subgroup $\Delta(12)$ \cite{hh99}.  This group is the same as $A_4$, 
the symmetry group of the even permutation of four objects.  It is also  
the symmetry group of the regular tetrahedron, one of five perfect 
geometric solids known to the ancient Greeks and identified by Plato with 
the element ``fire'' \cite{m02}.  There are four irreducible representations 
of $A_4$: 
\underline {1}, \underline {1}$'$, \underline {1}$''$, and \underline {3}, 
with the multiplication rule \cite{mr01}
\begin{equation}
\underline {3} \times \underline {3} = \underline {1} + \underline {1}' + 
\underline {1}'' + \underline {3} + \underline {3}.
\end{equation}
Under 
\begin{equation}
{\cal G} = SU(3)_C \times SU(3)_L \times SU(3)_M \times SU(3)_R \times A_4,
\end{equation}
the quark and lepton assignments are assumed to be
\begin{eqnarray}
&& q \sim (3,3,1,1;\underline{3}), ~~~ q^c \sim (3^*,1,1,3^*;\underline{3}), \\
&& l \sim (1,3^*,3^*,1;\underline{3}), ~~~ l^c \sim (1,1,3,3;\underline{3}),
\end{eqnarray}
with their electric charges given by
\begin{equation}
Q = I_{3L} + {Y_L \over 2} + I_{3R} + {Y_R \over 2} + I_{3M} - {Y_M \over 2}.
\end{equation}
A good visual summary of this scheme is Fig.~1.

\begin{figure}[htb]
\begin{center}
\begin{picture}(400,250)(0,0)
\ArrowLine(200,5)(150,55) 
\ArrowLine(250,55)(200,5)
\ArrowLine(250,55)(300,105)
\ArrowLine(300,105)(250,155)
\ArrowLine(200,205)(250,155)
\ArrowLine(150,155)(200,205)
\ArrowLine(150,155)(100,105)
\ArrowLine(100,105)(150,55)
\Text(200,214)[]{$SU(3)_C$}
\Text(200,-4)[]{$SU(3)_M$} 
\Text(79,105)[]{$SU(3)_L$} 
\Text(323,105)[]{$SU(3)_R$}
\Text(144,160)[]{$q$} 
\Text(259,160)[]{$q^c$} 
\Text(144,50)[]{$l$}
\Text(258,50)[]{$l^c$}
\Line(150,155)(180,125)
\Line(180,85)(150,55)
\Line(220,125)(250,155)
\Line(250,55)(220,85)
\Text(200,105)[]{$A_4$}
\end{picture}
\end{center}
\caption{Pictorial representation of three families of quarks and leptons.}
\end{figure}
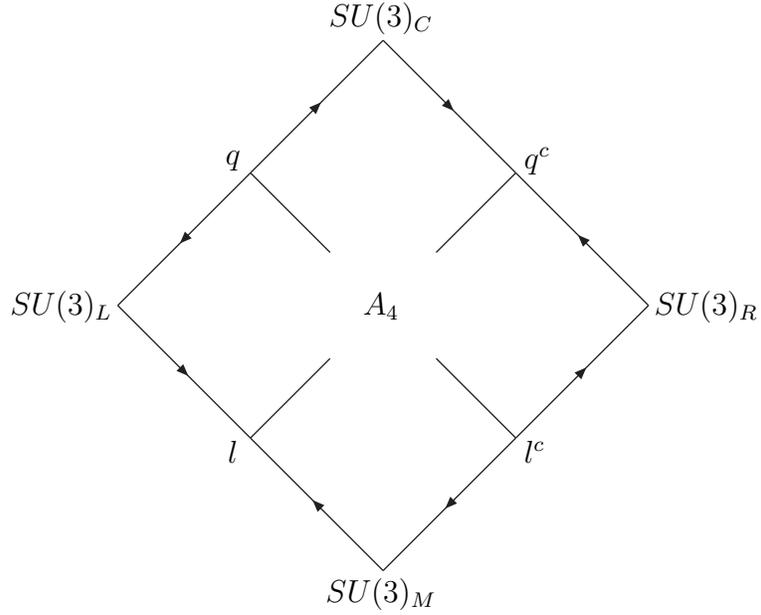

Given the fermionic content of Eqs.~(3) and (4), this theory is free of 
anomalies because each 3 is matched by a $3^*$ of the same multiplicity. 
There are also three interchange symmetries among the four $SU(3)$ 
groups, i.e.
\begin{eqnarray}
C \leftrightarrow M, ~~ L \leftrightarrow R &:& q \leftrightarrow l^c, ~~ 
q^c \leftrightarrow l; \\ 
C \leftrightarrow L, ~~ M \leftrightarrow R &:& q \leftrightarrow q, ~~ 
l^c \leftrightarrow l^c, ~~ l \leftrightarrow q^c; \\ 
C \leftrightarrow R, ~~ M \leftrightarrow L &:& l \leftrightarrow l, ~~ 
q^c \leftrightarrow q^c, ~~ q \leftrightarrow l^c.
\end{eqnarray}
These may be used to enforce the equality of the corresponding four gauge 
couplings, i.e. $g_C, g_L, g_M, g_R$, at the unification scale. 

Using the convention of previous models based on $[SU(3)]^3$ \cite {mmz04}, 
$[SU(3)]^4$ \cite {bmw04}, and $[SU(3)]^6$ \cite{m04}, where the rows have 
$I_3 = (1/2,-1/2,0)$ and the columns have $I_3 = (-1/2,1/2,0)$, the four 
matter multiplets are denoted as follows.
\begin{eqnarray}
SU(3)_C \times SU(3)_L &:& q = \pmatrix{d & u & h \cr d & u & h \cr 
d & u & h}, \\ 
SU(3)_R \times SU(3)_C &:& q^c = \pmatrix{d^c & d^c & d^c \cr 
u^c & u^c & u^c \cr h^c & h^c & h^c}, \\ 
SU(3)_L \times SU(3)_M &:& l = \pmatrix{N_1 & E^c_2 & \nu \cr 
E_1 & N_2 & e \cr N_3 & E^c_3 & S_1}, \\ 
SU(3)_M \times SU(3)_R &:& l^c = \pmatrix{N_4 & E^c_5 & N_6 \cr 
E_4 & N_5 & E_6 \cr \nu^c & e^c & S_2}.
\end{eqnarray}
In the above, the known quarks $(u,d,u^c,d^c)$ and leptons 
$(\nu,e,\nu^c,e^c)$ have their usual charges, i.e. $(2/3,-1/3,-2/3,1/3)$ 
and $(0,-1,0,1)$ respectively.  The exotic fields $(h,h^c,N,E,E^c,S)$  
have charges $(-1/3,1/3,0,-1,1,0)$ as given by Eq.~(5).  As shown below, 
the choice of \underline {3} of $A_4$ allows them all to be superheavy, 
whereas the more obvious choice of $3$ and $3^*$ of $SU(3)$ would not. 
Details on how the representations of $A_4$ are embedded in those of 
$SU(3)$ are given in the Appendix.

This model is now extended to include supersymmetry, and its low-energy 
particle content assumed to be that of the Minimal Supersymmetric 
Standard Model (MSSM).  Assuming the equality of all four gauge couplings 
at the unification scale, the value of $\sin^2 \theta_W$ from the 
contributions of $q$, $l^c$, $l$, and $q^c$ is given by
\begin{equation}
\sin^2 \theta_W = {\sum I_{3L}^2 \over \sum Q^2} = {{3 \over 2} + {3 \over 2} 
+ 0 + 0 \over 2 + 4 + 4 + 2} = {3 \over 12} = {1 \over 4},
\end{equation}
which is not equal to the desired values of $3/8$ for gauge-coupling 
unification \cite{bs04}.  However, if only the interchange symmetry of 
Eq.~(7) is used, and
\begin{equation}
g_M^2 = g_R^2 = 2 g_L^2 = 2 g_C^2
\end{equation}
is assumed, then
\begin{equation}
\sin^2 \theta_W = {{3 \over 2} + {3 \over 2} + 0 
+ 0 \over 2 + 3 + 2 + 1} = {3 \over 8}
\end{equation}
as desired.  The origin of Eq.~(14) is possibly the result of $SU(3)_C$ 
being the diagonal subgroup of $SU(3)_{CL} \times SU(3)_{CR}$, and $SU(3)_L$ 
that of $SU(3)_{qL} \times SU(3)_{lL}$, in which case $g_C^{-2} = g_{CL}^{-2} 
+ g_{CR}^{-2}$ and $g_L^{-2} = g_{qL}^{-2} + g_{lL}^{-2}$, so that Eq.~(14) 
is naturally obtained.

The quark and lepton multiplets of Eqs.~(3) and (4) are now supermultiplets 
together with two additional Higgs superfields
\begin{equation}
\Sigma \sim (1,3,1,3^*;\underline{1}), ~~~ \Sigma^c \sim (1,3^*,1,3;
\underline{1}).
\end{equation}
From Eq.~(5), it is obvious that both have the charge assignments of 
$l$ and $l^c$ of Eqs.~(11) and (12).  From the invariant terms $\Sigma 
\Sigma^c$ and $l l^c \Sigma$, it is also clear that the vacuum expectation 
values of the scalar fields $\langle \tilde \Sigma_{33} \rangle$, $\langle 
\tilde \Sigma^c_{33} \rangle$, $\langle \tilde S_1 \rangle$, and $\langle 
\tilde S_2 \rangle$ may all be naturally of order the unification scale.  
The symmetry breaking pattern is given by
\begin{eqnarray}
\langle \tilde \Sigma_{33} \rangle, \langle \tilde \Sigma^c_{33} \rangle 
\neq 0 &\Rightarrow& SU(3)_L \times SU(3)_R \to SU(2)_L \times SU(2)_R \times 
U(1)_{(Y_L+Y_R)/2}, \\
\langle \tilde S_1 \rangle \neq 0 &\Rightarrow& SU(3)_L \times SU(3)_M \to 
SU(2)_L \times SU(2)_M \times U(1)_{(Y_L-Y_M)/2}, \\
\langle \tilde S_2 \rangle \neq 0 &\Rightarrow& SU(3)_M \times SU(3)_R \to 
SU(2)_M \times SU(2)_R \times U(1)_{(Y_R-Y_M)/2},
\end{eqnarray}
resulting in
\begin{equation}
SU(3)_L \times SU(3)_M \times SU(3)_R \to SU(2)_L \times SU(2)_M \times 
SU(2)_R \times U(1)_{(Y_L+Y_R-Y_M)/2}.
\end{equation}

The fields which become superheavy at this stage are all the components of 
$\Sigma$ and $\Sigma^c$, as well as $h, h^c$ from the invariant 
$q^c q \Sigma^c$ term and $N_3, E^c_3, S_1, N_6, E_6, S_2$ from the 
invariant $l l^c \Sigma$ term.  To obtain the MSSM, the fields 
$N_1, E_1, E^c_2, N_2$ from $l$ and $N_4, E_4, E^c_5, N_5$ from $l^c$ 
must also become heavy.  Because of the assignments of Eq.~(4), the terms 
$l l l$ and $l^c l^c l^c$ are invariant from the product of three 
bitriplets under two $SU(3)$'s.  Thus $N_1 N_2 - E_1 E^c_2$ couples to 
$S_1$, and $N_4 N_5 - E_4 E^c_5$ couples to $S_2$, leaving only the 
particles of the MSSM (plus $\nu^c$) without any mass.  This is possible 
because both the symmetric and antisymmetric products of three 
\underline {3}'s in $A_4$ are singlets \cite{mr01}.  If $3$ and $3^*$ 
of $SU(3)$ are used instead, only the antisymmetric product is a singlet 
and that always leaves one zero eigenvalue in a $3 \times 3$ mass matrix.

The next task is to obtain the breaking
\begin{equation}
SU(2)_M \times SU(2)_R \times U(1)_{(Y_L+Y_R-Y_M)/2} \to U(1)_{Y/2}.
\end{equation}
Consider first $\langle \tilde N_5 \rangle \neq 0$.  This breaks $SU(2)_M 
\times SU(2)_R$ to $U(1)_{I_{3M}+I_{3R}}$, but does not break 
$U(1)_{(Y_L+Y_R-Y_M)/2}$.  At the same time, $\nu^c N_6 - N_4 S_2$ couples 
to $N_5$, so it appears at first sight that $\nu^c$ is now massive, but 
since $N_3 N_6 + E^c_3 E_6 + S_1 S_2$ couples to $\Sigma_{33}$, a linear 
combination of $\nu^c$ and $N_3$ will remain massless.  Neutrinos remain 
Dirac fermions in this case.  To obtain the desired breaking of Eq.~(21), 
$\langle \tilde N_6 \rangle \neq 0$ is also needed.  The $9 \times 9$ mass 
matrix spanning $\nu^c, N_3, N_4, N_5, N_6, S_1, S_2, \Sigma$, and $\Sigma^c$ 
must now be considered, and a careful analysis shows that $\nu^c$ gets a 
Majorana mass of order
\begin{equation}
m_{\nu^c} \sim {\langle \tilde N_5 \rangle^2 \langle \tilde N_6 \rangle^2 
\over M^3},
\end{equation}
where $M$ is the unification scale.  Note that this does not happen if either 
$\langle \tilde N_5 \rangle = 0$ or $\langle \tilde N_6 \rangle = 0$. 
The origin of this Majorana mass is the $4 \times 4$ mass submatrix 
spanning $S_1$, $S_2$, $\Sigma$, and $\Sigma^c$, due to the 
$S_1 S_2 \Sigma$ and $\Sigma \Sigma^c$ terms, i.e.
\begin{equation}
{\cal M}_{S\Sigma} = \pmatrix{0 & a & b & 0 \cr a & 0 & c & 0 \cr b & c & 0 
& d \cr 0 & 0 & d & 0},
\end{equation}
which is not of the Dirac form.  Suppose $\langle \tilde N_5 \rangle \sim 
\langle \tilde N_6 \rangle \sim 10^{-2} M$, then $m_{\nu^c} \sim 
10^{-8} M$, and for $M \sim 10^{16}$ GeV, $m_{\nu^c} \sim 10^{8}$ GeV 
which is very suitable for leptogenesis \cite{fy86,afs03}.  Note that 
the presence of $SU(3)_M$ in ${\cal G}$ of Eq.~(2) is necessary for 
obtaining this result.  Note also that in the well-known $SU(3)^3$ model 
with only triplets, it is impossible to get a small Majorana neutrino mass. 
In contrast, the quartic (instead of the canonical quadratic) 
seesaw formula of Eq.~(22) is automatic here in this model.

The last stage of symmetry breaking is
\begin{equation}
SU(2)_L \times U(1)_{Y/2} \to U(1)_Q
\end{equation}
for which the Higgs superfields
\begin{equation}
\Phi \sim (1,3,1,3^*;\underline{3}), ~~~ \Phi^c \sim (1,3^*,1,3;
\underline{3})
\end{equation}
are used, assuming nonzero values of $\langle \tilde \Phi_{11} \rangle$, 
$\langle \tilde \Phi_{22} \rangle$, $\langle \tilde \Phi^c_{11} \rangle$, 
$\langle \tilde \Phi^c_{22} \rangle$, which allow all leptons and quarks 
to become massive.  The invariant products $l l^c \Phi$ and $q^c q \Phi^c$ 
have both symmetric and antisymmetric terms under $\underline {3} \times 
\underline {3} \times \underline {3} \to \underline {1}$ in $A_4$, 
resulting in a $3 \times 3$ mass matrix for both $u$ and $d$ quarks of 
the form
\begin{equation}
{\cal M}_q = \pmatrix{0 & a & rc \cr ra & 0 & b \cr c & rb & 0}.
\end{equation}
Note that if ${\cal M}_q$ were antisymmetric, i.e. $r=-1$, then its 
eigenvalues would be zero and $\pm (|a|^2+|b|^2+|c|^2)^{-1/2}$, which is 
of course unrealistic.  On the other hand, if $r=0$, then its eigenvalues 
are simply $a$, $b$, and $c$.  Now Suppose $\Phi^c$ breaks along just one 
direction in family space, then the ratio $\langle \tilde \Phi^c_{11} 
\rangle /\langle \tilde \Phi^c_{22} \rangle$ is the same for each family, 
which means that ${\cal M}_u$ and ${\cal M}_d$ are proportional 
to each other.  They will thus undergo the same diagonalization (even if 
$r \neq 0$) and there will be no mixing.  This is exactly the situation 
in another model of grand unification considered previously \cite{bdm99} 
and the solution is to generate both the shifts in mass and the mixing 
matrix from the soft breaking of supersymmetry.  Note that $\Phi \Phi^c$, 
$\Phi \Phi \Phi$, $\Phi^c \Phi^c \Phi^c$, $\Phi \Phi \Sigma$, and $\Phi^c 
\Phi^c \Sigma^c$ are all allowed invariant terms under ${\cal G}$.  This 
means that there is no theoretical understanding of why the two Higgs 
doublets of the MSSM are light.  This problem is common to all models 
of grand unification and does not seem to have any special solution in 
the present context.

Consider now the lepton mass matrices.  They come from $l l^c \langle \tilde 
\Phi \rangle$, except that there are also large Majorana mass terms for 
$\nu^c$ already discussed.  The canonical seesaw mechanism \cite{seesaw,m05} 
applies in the usual way and very small Majorana neutrino masses are 
obtained.  Of course, important corrections from soft supersymmetry breaking 
are applicable in both quark \cite{bdm99} and lepton \cite{bmv03} sectors.

To summarize, it has been poroposed that the fundamental gauge theory of 
particle interactions is made up at the unification scale of four 
pairwise interchangeable $SU(3)$ factors, i.e. $SU(3)_C \times SU(3)_L 
\times SU(3)_M \times SU(3)_R$ plus one additional $SU(3)_F$ which 
determines the number of families.  Quark and lepton families belong to 
the \underline {3} representation of the discrete non-Abelian subgroup 
$A_4$ of $SU(3)_F$.  This structure has quark-lepton symmetry at high 
energies and yet allows naturally for the appearance of the MSSM at low 
energies.  It also allows neutrinos to obtain their small masses through 
the canonical seesaw mechanism. Soft supersymmetry breaking terms which 
also break $A_4$ are necessary, although the details cannot be uniquely 
determined.  In other words, whereas the particle content of this model 
is the same as that of the MSSM, its soft supersymmetry breaking sector 
has to be very different and that is something which experiments in the 
near future can confirm or disprove.

This work was supported in part by the U.~S.~Department of Energy under 
Grant No. DE-FG03-94ER40837.

\newpage
\bibliographystyle{unsrt}

\newpage
\noindent \underline {Appendix}.  The well-known low-dimensional 
representations of $SU(3)$ transform under $A_4$ as follows:
\begin{eqnarray}
3, ~ 3^* &\sim& \underline {3}, \\ 
6, ~ 6^* &\sim& \underline {1} + \underline {1}' + \underline {1}'' + 
\underline {3}, \\ 
8 &\sim& \underline {1}' + \underline {1}'' + \underline {3} + 
\underline {3}, \\ 
10, ~ 10^* &\sim& \underline {1} + \underline {3} + \underline {3} + 
\underline {3}.
\end{eqnarray}
Graphically, the \underline {3} of $A_4$ may be represented by a triangle. 
There are one, two, and three such (dashed) triangles in the 6, 8, and 10 
representations of $SU(3)$ respectively as depicted below.

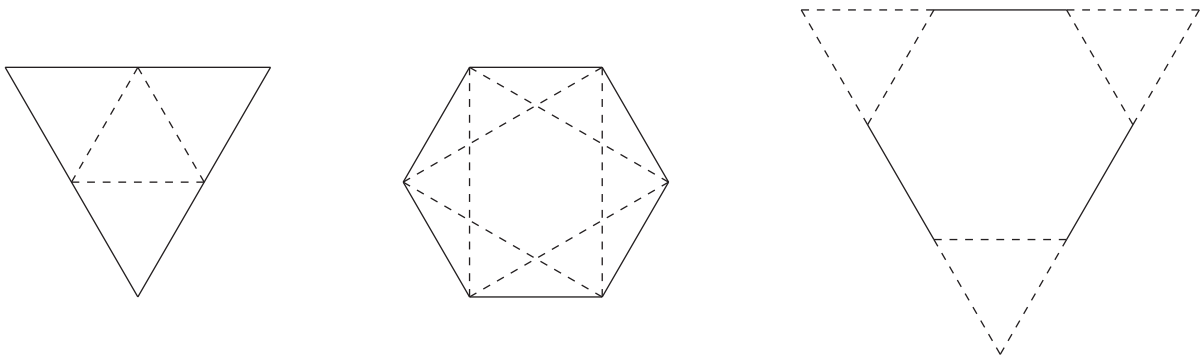
\begin{figure}[htb]
\begin{center}
\begin{picture}(450,150)(0,0)
\Line(0,108.25)(50,21.65)
\Line(50,21.65)(100,108.25)
\Line(0,108.25)(100,108.25)
\DashLine(25,64.95)(75,64.95)3
\DashLine(25,64.95)(50,108.25)3
\DashLine(50,108.25)(75,64.95)3

\Line(150,64.95)(175,21.65)
\Line(150,64.95)(175,108.25)
\Line(175,108.25)(225,108.25)
\Line(175,21.65)(225,21.65)
\Line(225,108.25)(250,64.95)
\Line(225,21.65)(250,64.95)
\DashLine(175,21.65)(175,108.25)3
\DashLine(225,21.65)(225,108.25)3
\DashLine(175,108.25)(250,64.95)3
\DashLine(175,21.65)(250,64.95)3
\DashLine(150,64.95)(225,108.25)3
\DashLine(150,64.95)(225,21.65)3

\DashLine(300,129.9)(350,129.9)3
\Line(350,129.9)(400,129.9)
\DashLine(400,129.9)(450,129.9)3
\DashLine(300,129.9)(325,86.6)3
\Line(325,86.6)(350,43.3)
\DashLine(350,43.3)(375,0)3
\DashLine(350,43.3)(400,43.3)3
\DashLine(375,0)(400,43.3)3
\Line(400,43.3)(425,86.6)
\DashLine(425,86.6)(450,129.9)3
\DashLine(425,86.6)(400,129.9)3
\DashLine(325,86.6)(350,129.9)3

\end{picture}
\end{center}
\caption{Pictorial representations of the \underline {3} of $A_4$ in $SU(3)$.}
\end{figure}

\end{document}